\documentclass[letterpaper, 10 pt, conference]{style/ieeeconf}  % Comment this line out if you need a4paper

\IEEEoverridecommandlockouts                              % This command is only needed if 
\usepackage{graphics} % for pdf, bitmapped graphics files
\usepackage{epsfig} % for postscript graphics files
\usepackage{amsmath} % assumes amsmath package installed
\usepackage{multirow}
\usepackage{bm}
\usepackage{color}
\usepackage{tabularx}
\usepackage{subfigure}
\usepackage{enumerate}

\title{\LARGE \bf
Effect of Haptic Feedback on Avoidance Behavior and Visual Exploration in Dynamic VR Pedestrian Environment
}

\author{Kyosuke Ishibashi$^{1}$, Atsushi Saito$^{1}$, Zin Y. Tun$^{2}$, Lucas Ray$^{3}$, Megan C. Coram$^{3}$, \\ Akihiro Sakurai$^{1}$, Allison M. Okamura$^{3}$, and Ko Yamamoto$^{1}$% <-this % stops a space
\thanks{*This work is supported by JST PRESTO Grant Number JPMJPR226A}% <-this % stops a space
\thanks{$^{1}$Kyosuke Ishibashi, Atsushi Saito, Akihiko Sakurai and Ko Yamamoto are with Department of Mechano-informatics, The University of Tokyo, 7-3-1 Hongo, Bunkyo-ku, Tokyo, 113-8656, Japan.
        {\tt\small ishibashi-kyosuke@ynl.t.u-tokyo.ac.jp}}%
\thanks{$^{2}$Zin Y. Tun is with the Department of Electrical Engineering, Stanford University, Stanford, CA, 94305, USA.}%
\thanks{$^{3}$Lucas Ray, Megan C. Coram and Allison M. Okamura are with the Department of Mechanical Engineering, Stanford University, Stanford, CA, 94305, USA.}%
}

\begin{document}

\maketitle
\thispagestyle{empty}
\pagestyle{empty}

%%%%%%%%%%%%%%%%%%%%%%%%%%%%%%%%%%%%%%%%%%%%%%%%%%%%%%%%%%%%%%%%%%%%%%%%%%%%%%%%
\begin{abstract}

Human crowd simulation in virtual reality (VR) is a powerful tool with potential applications including emergency evacuation training and assessment of building layout. While haptic feedback in VR enhances immersive experience, its effect on walking behavior in dense and dynamic pedestrian flows is unknown. Through a user study, we investigated how haptic feedback changes user walking motion in crowded pedestrian flows in VR. The results indicate that haptic feedback changed users' collision avoidance movements, as measured by increased walking trajectory length and change in pelvis angle. The displacements of users' lateral position and pelvis angle were also increased in the instantaneous response to a collision with a non-player character (NPC), even when the NPC was inside the field of view. Haptic feedback also enhanced users' awareness and visual exploration when an NPC approached from the side and back. Furthermore, variation in walking speed was increased by the haptic feedback.
These results suggested that the haptic feedback enhanced users' sensitivity to a collision in VR environment.

\end{abstract}

%%%%%%%%%%%%%%%%%%%%%%%%%%%%%%%%%%%%%%%%%%%%%%%%%%%%%%%%%%%%%%%%%%%%%%%%%%%%%%%%
\section{INTRODUCTION}

Human crowd simulation is important for navigation in daily life, design of a building layout, evacuation guidance in an emergency situation, and evaluation of crowd crush risks.
Failure in predicting congestion can lead to a serious crowd crush accidents, e.g., Akashi Pedestrian Bridge Accident (2001, Japan), Love Parade disaster (2010, Germany) \cite{Zhao2020}, and Itaewon crowd crush (2022, Korea).
Those accidents happened because people became crowded in a space due to high input flow rate and limited egress.
Once people are crowded, it becomes hard to escape from the dense area.
To prevent such a serious accident, it is important to predict how much congestion will occur and appropriately guide people before they are crowded.

\begin{figure}[t]
    \centering
    \subfigure[Real environment]{
        \includegraphics[width=0.44\hsize]{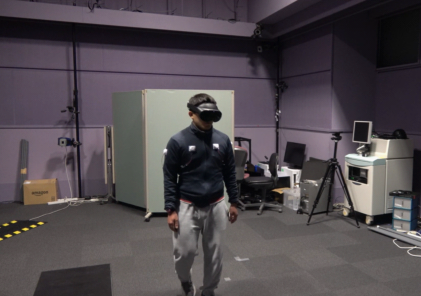}}
    \subfigure[VR environment]{
        \includegraphics[width=0.46\hsize]{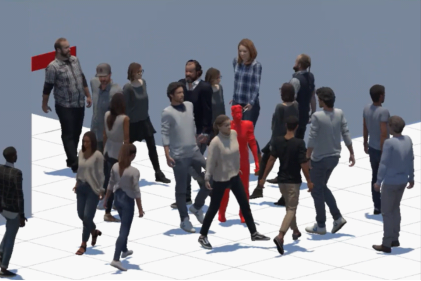}}
    \caption{\small{Experimental scene of a user study for haptic feedback in VR pedestrian flow simulation: (a) real environment and (b) VR environment. A red character in (b)  indicates the subject motion in the VR simulation.}}
    \label{fig:VRexperiment}
\end{figure}

In the literature, there are many studies of pedestrian crowd simulation, including multi-agent \cite{Helbing1995}, continuum \cite{henderson_1971_nature,Treuille2006,Yamamoto2013RAS}, and cellular automata \cite{Blue2000,Asano2006,Chen2018} models.
In multi-agent simulation, one approach to generate a pedestrian behavior is rule-based one.
One representative method is Social Force Model \cite{Helbing1995}.
In particular, several methods were proposed  for local collision avoidance behavior \cite{Berg2008}.
With advances in machine learning techniques, data-driven approaches have been also used for crowd simulation.
Many approaches use a long short-term memory (LSTM)-type recurrent neural network, e.g., Social LSTM \cite{Alahi2016} and Social-Scene-LSTM \cite{Xue2018}, and attention mechanism \cite{Vemula2018ICRA,Sakurai2024ROMAN}.

An integration of such a human crowd simulation with virtual reality \cite{Ulicny2001,Olivier2014,Nelson2019} is a powerful tool with potential applications including evacuation training \cite{Zhang2023,Zhang2023b} and assessment of building layout \cite{Ventura2018}.
In a VR system, haptic feedback provides more immersive experience for a user.
There are several studies on how user behavior changes with haptic feedback in a human crowd VR simulation. Table \ref{tab:features_of_studies} summarizes the features of these studies.
Venkatesan et al. \cite{Venkatesan2023} conducted a user experiment using a projection-based VR system and a haptic feedback on the shoulder.
Krogmeier et al. \cite{Krogmeier2019} investigated users' emotional arousal by measuring galvanic skin response assuming a situation where a user is standing still wearing a head-mounted display (HMD) and a haptic vest while non-player characters (NPC) walk past them.

\begin{table*}[!t]
    \centering
    \small
    \begin{tabularx}{\textwidth}{X|lXll}
       \textbf{Related works} & \textbf{Density of NPCs} & \textbf{Pedestrian flow setting} & \textbf{User's behavior} & \textbf{Haptic feedback} \\ \hline
       Venkatesan et al. \cite{Venkatesan2023} & Low/High & Opposite to User & Standing & Arms \\
       Krogmeier et al. \cite{Krogmeier2019} & One at a Time & Four Directions & Standing & Torso \\
       Koilias et al. \cite{Koilias2020} & High & Same as User & Walking & Torso \\
       Berton et al. \cite{Berton2022} & High & Stationary & Walking & Arms \\
       Yun et al. \cite{Yun2024} & Low & Opposite to User/Four Directions & Walking & Arms \\
       \hline
       Our study & High & Four Intersecting Flows & Walking & Arms and Torso \\
    \end{tabularx}
    \caption{Comparison of experimental setup among related works and this study.}
    \label{tab:features_of_studies}
\end{table*}

Koilias et al. \cite{Koilias2020} also used a vest-type haptic suit and investigated walking behavior in a single-directional pedestrian flow.
Berton et al. \cite{Berton2022} quantitatively evaluated users' walking trajectories with haptic feedback on the arms assuming the NPCs were standing still.
Yun et al. \cite{Yun2024} investigated the effect of haptic feedback on the arms in a four-directional crossing flows, although each flow was relatively sparse. 
In those studies, it was reported that haptic feedback improved realistic sensation, but other studies \cite{Berton2022} reported that haptic feedback did not significantly affect walking trajectories.
Moreover, prior studies only dealt with a situation where a user is or NPCs are standing still, a single pedestrian flow, or multi-directional but sparse pedestrian flows, but did not consider more {\it dynamic} situation in which a user walks through multi-directional and dense flows.

In this study, we investigate the effect of haptic feedback in such a dynamic pedestrian flows.
It was reported that a human locomotion behavior mainly depends on visual perception \cite{Patla1997}, and Berton et al. \cite{Berton2022} considered that was a reason why a haptic feedback did not affect user's walking path selection.
However, it is expected that the visual information is limited in a dynamic situation because a number of NPCs are approaching from different directions, in addition to the limitation of the field of view of an HMD.
%Therefore, we establish our research question:
%\begin{itemize}
%    \item How does a haptic feedback affect a user's walking trajectory in a VR situation where NPCs are walking following four-directional flows?
%\end{itemize}
We fabricated a haptic suit in which six vibration motors are attached for omni-directional perception, and conducted a user study for pedestrian flow in a VR simulation.

\section{Human Pedestrian Flow VR Simulator}
\label{chap:simulation}

\subsection{Pedestrian Trajectory Model}
We used a VR pedestrian flow simulator created by Sakurai et al. \cite{sakurai2023vr}. In this simulator, NPCs move based on the pedestrian model by Yamamoto and Okada \cite{yamamoto2013control}, where each NPC's trajectory is generated by a rule-based algorithm similar to the Social Force Model \cite{Helbing1995}.
The 2D walking velocity vector is given as the sum of two types of effects.
One is a global navigation velocity given by a 2D velocity vector field \cite{yamamoto2013control}, in which we can set a parameter for {\it base} velocity that is an ideal walking speed if there is only a single pedestrian.
The other is a repulsive velocity for the collision avoidance with the other NPCs (including a VR user). 
The walking trajectory is obtained by integrating the walking velocity.
This kind of velocity-based modeling allows us to easily set an overall walking velocity for different experimental conditions.

\subsection{Motion Matching}
A full-body NPC motion is generated from the 2D walking trajectory by the motion matching algorithm proposed by B{\"u}ttner and Clavet \cite{buttner2015motion}. The motion matching algorithm uses two databases: a matching database consisting of feature vectors $\bm{x}$ representing character states, and an animation database consisting of pose vectors $\bm{y}$ representing character motions. We compute a search vector $\bm{q}$ that represents the character current state and use the matching database to determine a suitable frame $f$ based on the distance between $\bm{q}$ and $\bm{x}$.
\newcommand{\argmin}{\mathop{\rm arg~min}\limits}
\begin{align}
    f = \argmin_k \| \bm{q} - \bm{x}_k \|   
\end{align}
By searching the animation database for the pose vector corresponding to frame $f$, we can efficiently obtain the character's posture. 

To reduce the computational cost and enable real-time rendering, motion matching is omitted for NPCs outside the HMD's view, keeping animations realistic for visible NPCs \cite{sakurai2023vr}.

\section{Experimental Design}
\label{chap:design}

\subsection{Hardware}

\begin{figure}[t]
    \centering
    \subfigure[Side view]{
        \includegraphics[width=0.35\hsize]{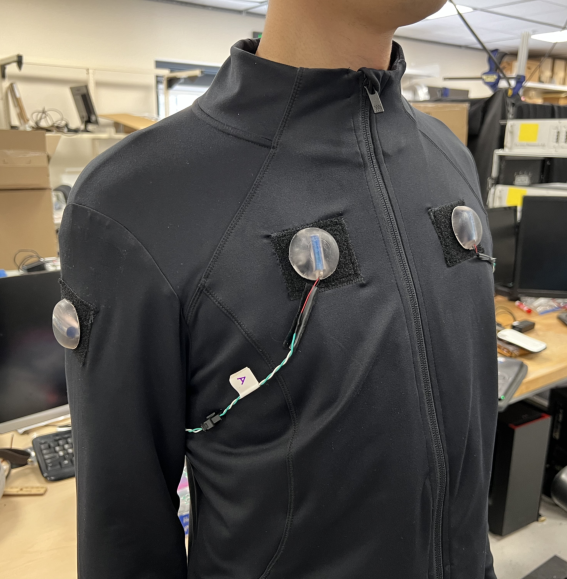}}
    \subfigure[Top view]{
        \includegraphics[width=0.35\hsize]{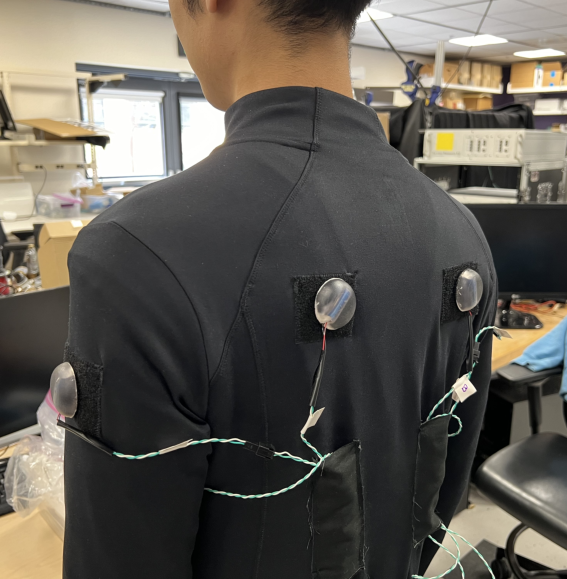}}
    \subfigure[Haptic Unit]{
        \includegraphics[width=0.3\hsize]{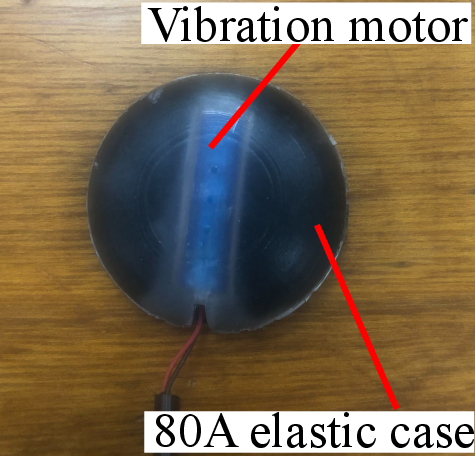}}
    \subfigure[Haptic Rendering Pattern]{
        \includegraphics[width=0.4\hsize]{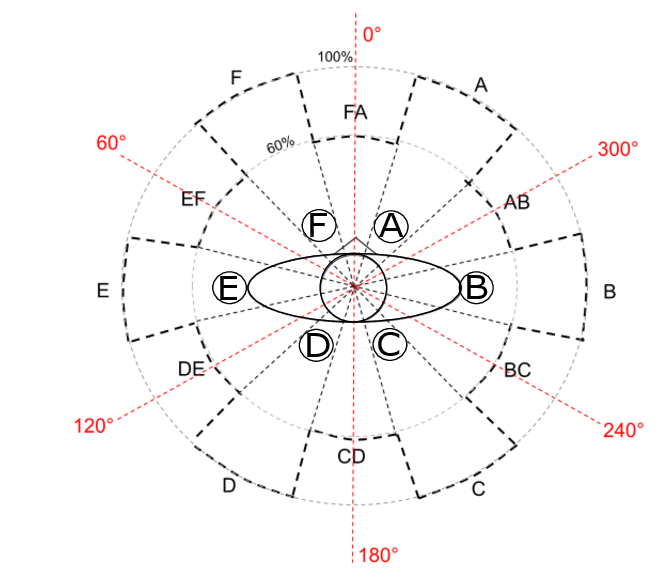}}
    \caption{\small{Haptic wearable suit to render haptic collisions between subjects and virtual characters.  These units are attached to the left and right sides of the chest, upper arms, and upper back.}}
    \label{fig:haptic_suit}
\end{figure}

\subsubsection{Computer}
We used ZEFT PC (specification: Core i9-14900K, MSI RTX4090 Gaming X Slim, 128GB) for calculating the pedestrian trajectory model and motion matching.

\subsubsection{Head Mounted Display (HMD)}
We used Meta Quest2 to immerse participants in the virtual environment (specifications: 120 Hz, $120^{\circ}$FoV, 1832×1920 resolution).

\subsubsection{Motion Capture}
To provide a self-avatar in VR, we used Mocopi \cite{mocopi}, which is an IMU-based (Inertial Measurement Unit) motion capture system. Mocopi estimates human joint angles by six small sensors attached to the head, wrists, ankles, and hip. 

%Note that the precision of Mocopi is not as high as another motion capture suit such as Xsens.
During the experiment, a user's self avatar was visualized for immersive experience using the joint angle data obtained by Mocopi.
Because the position of the head obtained by Mocopi also included an error, we rendered the avatar's head position using measurements from the HMD.  

\subsubsection{Haptic Device}
To render haptic collisions between subjects and virtual characters, we developed a haptic wearable suit, as shown in Fig. \ref{fig:haptic_suit}. Each haptic unit consists of a vibration motor and a 3D-printed case made from Shore A 80 rubberlike material. These units are attached to the left and right sides of the chest, upper arms, and upper back. This arrangement can provide the omni-directional information about collision to a user whereas most of the related works provided haptic feedback to either the shoulders or the chest/back, as summarized in Table \ref{tab:features_of_studies}. Based on the direction of the colliding NPC, the haptic suit can generate 12 patterns of feedback in 30-degree increments by activating up to two motors per collision. We use Bluetooth communication with an ESP32 to actuate the haptic units.

\subsection{VR Environment}
\begin{figure}[t]
    \centering
    \includegraphics[width=0.7\hsize]{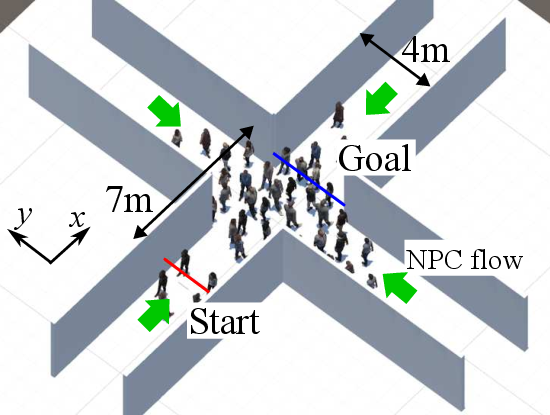}
    \caption{\small{Bird's-eye view of the VR environment. In the experiment, a subject was instructed to walk in a crossing hallway from the start position indicated by the red bar to the goal position indicated by the blue bar, avoiding NPCs.}}
    \label{fig:bird_view}
\end{figure}

%Fig. \ref{fig:VRexperiment} and 
Fig. \ref{fig:bird_view} shows a bird's-eye view of the VR environment used in the user study. In each trial of the experiment, a subject was instructed to walk in a crossing hallway from the start position indicated by the red bar to the goal position indicated by the blue bar, avoiding NPCs.
The width of the hallway is 4 m, and the distance between the start and goal positions is 7 m.
There were four pedestrian flows, shown by the green arrows, in the $\pm x$ and $\pm y$ directions.
NPCs in each flow were generated by the method outlined in Section \ref{chap:simulation}.
%Four types of NPCs are generated in the environment: one walking in the same direction as the subject (+x), one walking in the opposite direction (-x), and others walking toward the left and right in the intersection (+y and -y). Their movements are generated based on the algorithm described in Section \ref{chap:simulation}. 
The base walking speed of each velocity vector field was set to 1.2 m/s, and each NPC was generated every 1.3 seconds. 
These values were determined from the result of a preliminary experiment.

Fig. \ref{fig:VRexperiment} shows an experimental scene of the user study in (a) real and (b) VR environments.
The red character in (b) indicates the subject motion measured by Mocopi and HMD.
When the distance between a subject and other NPCs becomes less than 0.65 m, haptic feedback is triggered depending on the collision direction. We use the Mega Bundle Casual Rigged 001 – 005 from RenderPeople Inc. as NPC appearance datasets.

\subsection{Experimental Procedure}
We conducted the experiment with twenty unpaid subjects (6 female and 14 male) at two locations: 9 at the Stanford Robotics Center at Stanford University and the other 11 at Cyber Behavior Studio in the Department of Mechano-informatics, the University of Tokyo. 
Ethics approval for the experimental procedures was granted by Stanford University (protocol \#22514: “Haptics in Virtual Environment and Teleoperation Systems,”) (to be checked), and the Human Research Committee at Graduation School of Information Science and Technology, the University of Tokyo (approval No.: 22-408).

%Nine subjects participated at Stanford University, while eleven took part at the University of Tokyo. 
The subjects were divided into two groups.
In both groups, 10 trials run for each subject.
{\hyphenpenalty=10000
\begin{itemize}
    \item In {\bf Group 1}, the first session consisted of of 1 practice followed by 5 trials without haptic feedback (\textit{NoHaptics}); and the last session consisted of 1 practice followed by 5 trials with haptic feedback (\textit{Haptics}).
    \item In {\bf Group 2}, the first session was 1 practice followed by 5 trials of \textit{Haptics}; and the last session was 1 practice followed by 5 trials of \textit{NoHaptics}.
\end{itemize}
}
%The experiment consisted of two blocks: one with the haptic device and one without it. Each block included one practice trial followed by five experimental trials. To counterbalance the order of conditions, subjects were divided into two groups: Group 1 first walked through the environment without the haptic device and then with it, while Group 2 experienced the conditions in the opposite order. 
After each session, a subject answered  a questionnaire listed in Table \ref{tb:questionnaire} for qualitative evaluation.

\subsection{Hypotheses and Measurements}
%These are our hypotheses of the experiment.
\begin{enumerate}[H1)]
    \item Haptic feedback does not change subjects' walking behavior including collision avoidance as a human locomotion mainly depends on visual perception \cite{Patla1997}.
    \item Haptic feedback increases subjects' visual exploration as he/she can notice an NPC approaching from outside of the field of view.
    \item Haptic feedback decreases subjects' walking speed as he/she tries to walk more carefully \cite{Koilias2020}. 
\end{enumerate}

For H1, we analyzed the walking trajectory and pelvis angle from the goal direction (Section \ref{chap:discussion}-A). For H2, we analyzed head angle with respect to the pelvis (Section \ref{chap:discussion}-B). For H3, we calculated the average walking speed of the subjects and its standard deviation (Section \ref{chap:discussion}-C). Additionally, when a subject collided with a NPC, we analyzed their movements immediately after the collision.

\begin{figure*}[t]
    \centering
    \subfigure[Trajectory length]{
        \includegraphics[width=0.25\hsize]{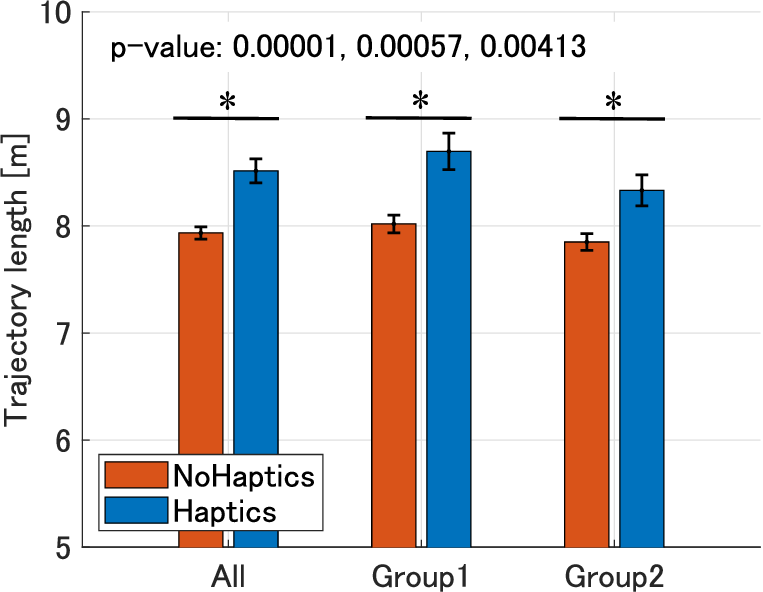}}
    \subfigure[Average pelvis angle toward goal direction]{
        \includegraphics[width=0.25\hsize]{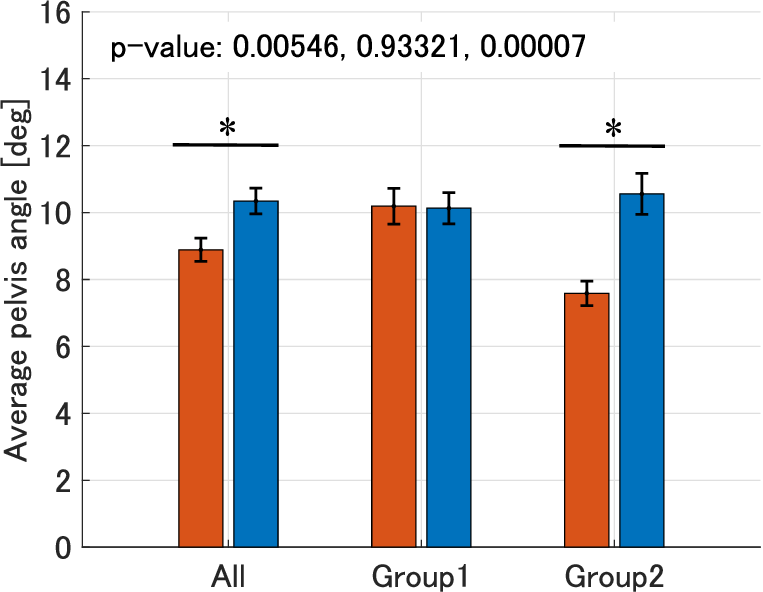}}
    \subfigure[Average head angle relative to pelvis angle]{
        \includegraphics[width=0.25\hsize]{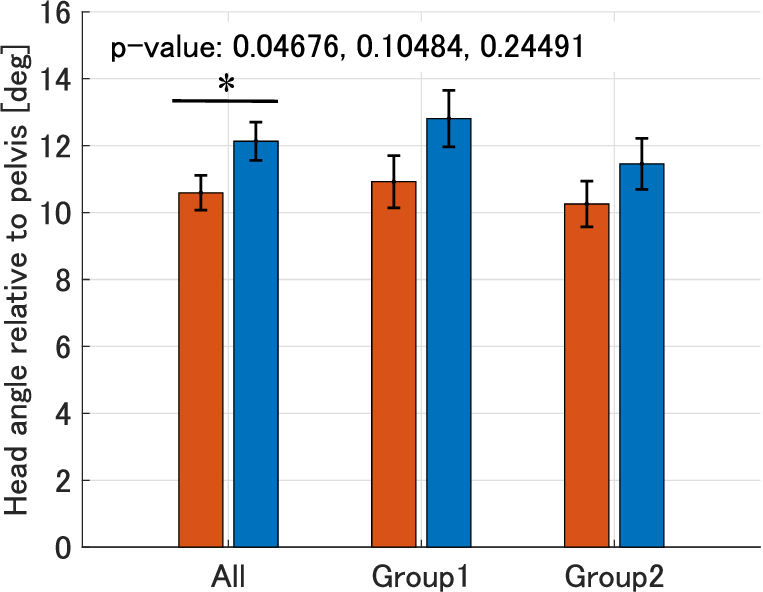}}
    \subfigure[Average walking speed]{
        \includegraphics[width=0.25\hsize]{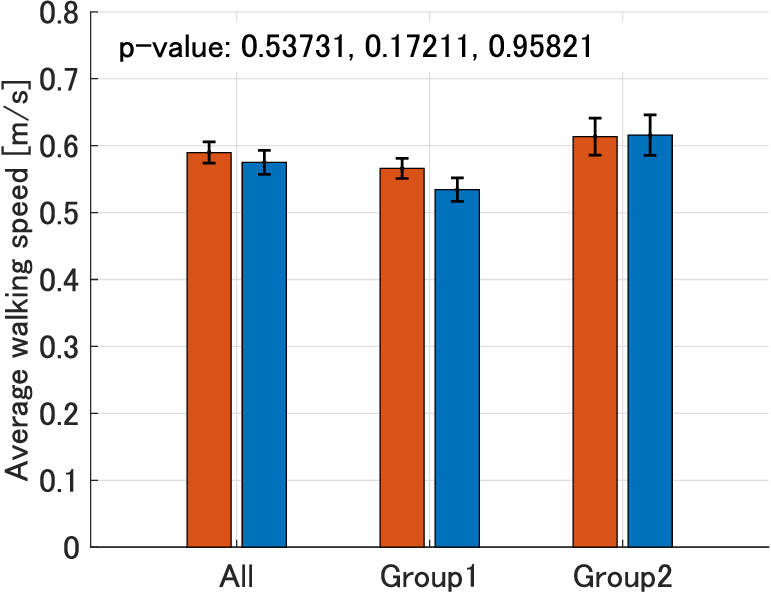}}
    \subfigure[Standard deviation of walking speed]{
        \includegraphics[width=0.25\hsize]{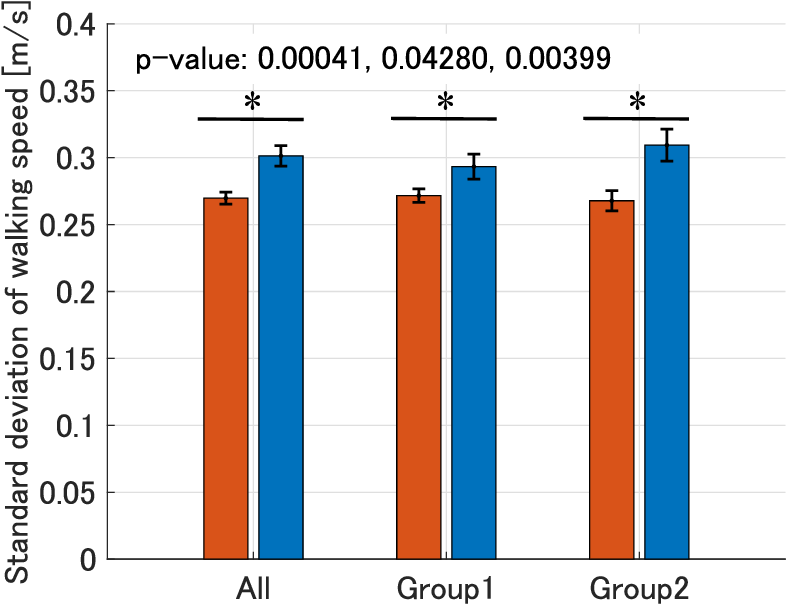}}
    \subfigure[Number of collision]{
        \includegraphics[width=0.25\hsize]{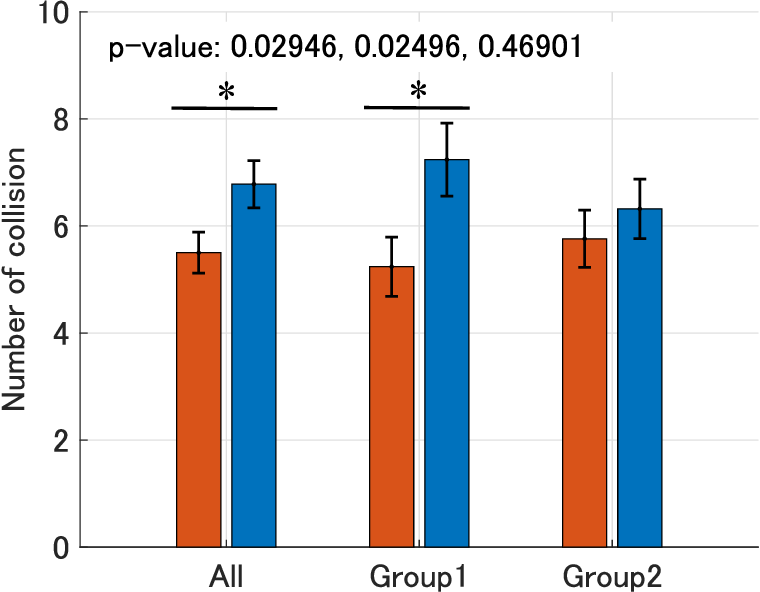}}
    \caption{\small{Walking motion data of VR experiments. The blue bars represent data from trials \textit{Haptics}, while the red bars show data from trials \textit{NoHaptics}. In each graph, the left bar group presents data for all subjects, the center bar group displays data for subjects in Group 1 (as described in Section \ref{chap:design}-C), and the right bar group shows data for subjects in Group 2. Error bars indicate standard error, and the p-values for the significant of haptic condition are shown at the top left of each graph, and horizontal lines are shown above groups with significant differences between Haptics and No Haptics, with a threshold of $p<0.05$.}}
    \label{fig:walking_motion}
\end{figure*}
\section{results}
\label{chap:results}

\subsection{Average Walking Behavior}
To verify the hypotheses, we calculated the average values of (a) the walking trajectory length, (b) pelvis angle displacement toward the goal direction, (c) head angle relative to pelvis angle, (d) walking speed and (e) its standard deviation, and (f) the number of collisions of a subject. Fig. \ref{fig:walking_motion} shows the results of these metrics, in which the red and blue bars indicate the results in \textit{NoHaptics} and \textit{Haptics} cases, respectively. In each graph, the left, middle and right sets of the bars indicate the average values of all subjects, Group 1, and Group 2, respectively. Error bars indicate the standard error, and the p-values are shown at the top left of each graph.

Note that some of related works \cite{Berton2022} employed the shoulder rotation angle as a metric of avoidance movement.
In this study, we employed the pelvis rotation angle because the shoulder angle obtained from Mocopi had an error.

\subsection{Instantaneous Response of Collision Avoidance and Visual Exploration}

\begin{figure*}[t]
    \centering
    \subfigure[Forward displacement]{
        \includegraphics[width=0.23\hsize]{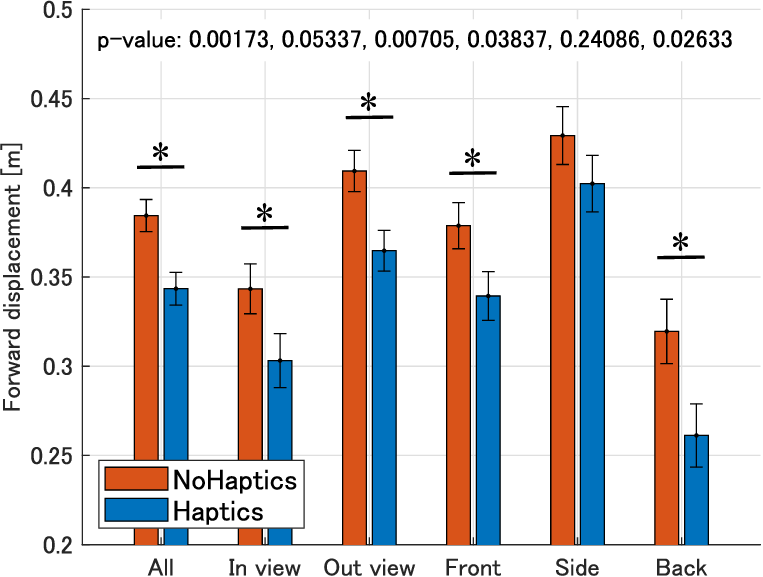}}
    \subfigure[Absolute value of lateral displacement]{
        \includegraphics[width=0.23\hsize]{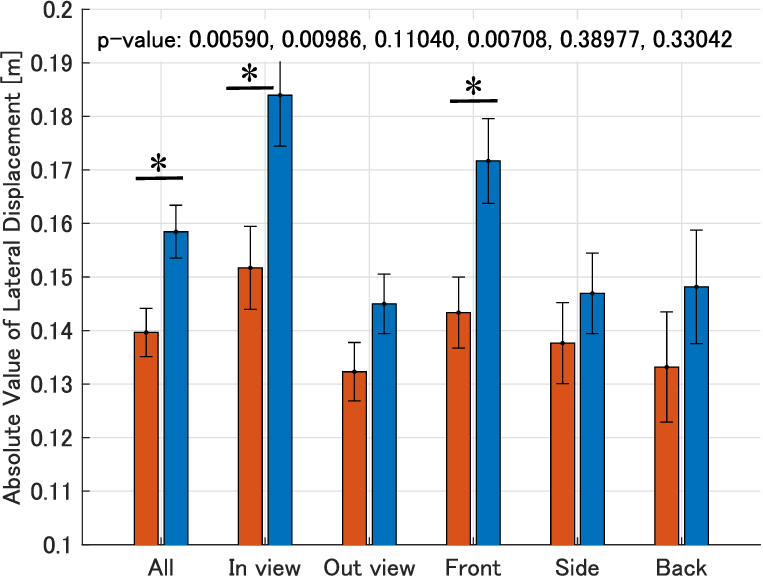}}
    \subfigure[Max. displacement of pelvis angle]{
        \includegraphics[width=0.23\hsize]{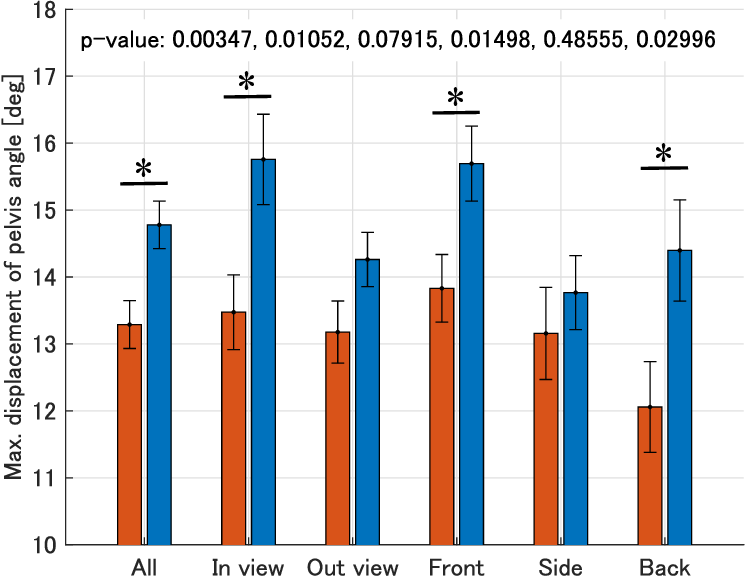}}
    \subfigure[Max. displacement of head angle from pelvis]{
        \includegraphics[width=0.23\hsize]{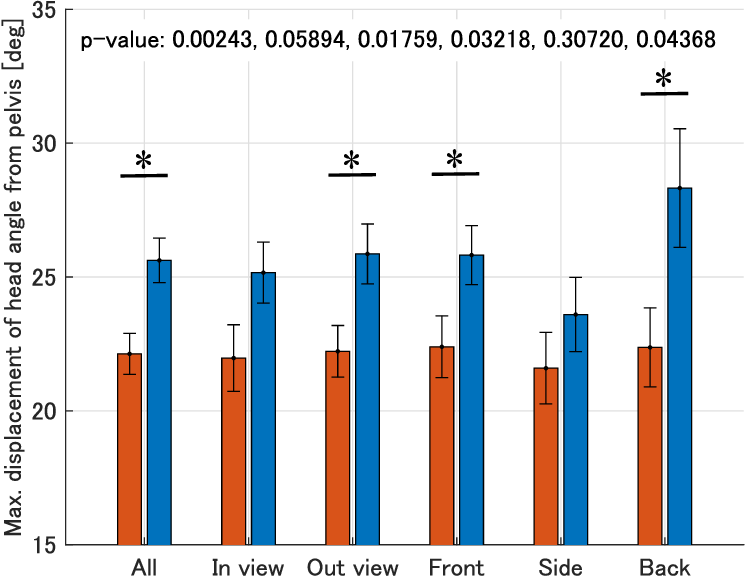}}
    \caption{\small{Collision avoidance motion: (a) the forward and (b) lateral displacements, the displacements of (c) the pelvis and (d) head angles after a collision. For head and pelvis angles, we analyze the maximum head or pelvis angle during the one-second period following the collision. Error bars indicate standard error, and the p-values for the significant of haptic condition are shown at the top left of each graph, and horizonize lines are shown above groups with significant differences between Haptics and NoHaptics, with a threshold of $p<0.05$.}}
    \label{fig:collision_motion}
\end{figure*}

To further explore H1 and H2, we analyzed how a subject responded immediately after colliding with NPCs, focusing on the data in one second after the collision. Fig. \ref{fig:collision_motion} shows (a) the forward and (b) lateral displacements, the displacements of (c) the pelvis and (d) head angles after a collision. 

For (a) the forward displacement, we calculated the displacement of the subject's position in the $x$-axis at one second after a collision from that at the moment of the collision, and then plotted its average across collisions. For (b) the lateral displacement, we calculated the absolute value of the displacement in the $y$-axis and its average.
For (c) the pelvis and (b) head angles, we extracted the maximum displacement during the one-second period after the collision and calculated its average. 
Note that the value of the head angle is the relative angle from the pelvis to HMD to represent the neck rotation angle for visual exploration.

In each graph, we also calculated the average value for the following five cases of the collisions: the two cases that an NPC was approaching from the inside or outside of the field of view ({\it In-view} and {\it Out-view}), and three cases that an NPC was approaching from the front, side or back ({\it Front}, {\it Side}, or {\it Back}). 

\subsection{Self-reported Ratings}
%To support the results obtained from subject's walking motion, 
We also qualitatively evaluated subjects' experience in the VR simulation by a questionnaire with a 5-point Likert scale (1 = strongly disagree, 5 = strongly agree). There were total 12 questions divided into four categories: Presence, Embodiment, Reality, and Awareness. 

Table \ref{tb:questionnaire} presents the question sets and scores for conditions \textit{Haptics} and \textit{NoHaptics}. The results showing a significant difference (\textit{p} $<$ 0.05) are highlighted in bold in the table.
\begin{table*}[]
    \centering
    \begin{tabular}{llcc}
        Categories & Questions & {\it NoHaptics} & {\it Haptics} \rule[-7pt]{0pt}{14pt} \\ \hline \hline
        \multirow{3}{*}{\textbf{Presence}} & Q1: I felt like I was in a real crowded space & 3.7 $\pm$ 0.2 & 4.1 $\pm$ 0.1 \rule[-4pt]{0pt}{12pt} \\
        & Q2: I felt stressed while walking through the environment & 3.8 $\pm$ 0.2 & 4.2 $\pm$ 0.2 \rule[-4pt]{0pt}{12pt} \\
        & \textbf{Q3: I was worried about colliding with NPCs} & \textbf{3.3 $\pm$ 0.3} & \textbf{4.2 $\pm$ 0.1}  \rule[-4pt]{0pt}{12pt} \\ \hline
        \multirow{3}{*}{\textbf{Embodiment}} & Q4: The movements of the virtual body felt like my own & 3.9 $\pm$ 0.2 & 4.1 $\pm$ 0.2 \rule[-4pt]{0pt}{12pt} \\
        & Q5: I felt like I was in control of the virtual body's movements & 4.0 $\pm$ 0.2 & 4.3 $\pm$ 0.2 \rule[-4pt]{0pt}{12pt} \\
        & Q6: The movements of the virtual body were synchronized with my own movements & 4.2 $\pm$ 0.2 & 4.6 $\pm$ 0.1 \rule[-4pt]{0pt}{12pt} \\ \hline 
        \multirow{3}{*}{\textbf{Reality}} & Q7: I think my movements were similar to how I would behave in a real situation & 3.5 $\pm$ 0.2 & 4.0 $\pm$ 0.2 \rule[-4pt]{0pt}{12pt} \\
        & Q8: I felt like the NPCs were actively trying to avoid me & 2.3 $\pm$ 0.2 & 2.2 $\pm$ 0.2 \rule[-4pt]{0pt}{12pt} \\
        & Q9: I was able to walk in sync with the other NPCs & 3.1 $\pm$ 0.2 & 3.1 $\pm$ 0.2 \rule[-4pt]{0pt}{12pt} \\ \hline
        \multirow{3}{*}{\textbf{Awareness}} & \textbf{Q10: I noticed NPCs approaching from the sides or behind me} & \textbf{2.9 $\pm$ 0.3} & \textbf{4.0 $\pm$ 0.3} \rule[-4pt]{0pt}{12pt} \\
        & \textbf{Q11: I paid attention to NPCs coming from the sides or behind me} & \textbf{3.0 $\pm$ 0.2} & \textbf{3.9 $\pm$ 0.2} \rule[-4pt]{0pt}{12pt} \\
        & Q12: I think I was able to successfully avoid collisions with NPCs & 2.5 $\pm$ 0.2 & 2.9 $\pm$ 0.2 \rule[-4pt]{0pt}{12pt} \\ \hline
    \end{tabular}
    \caption{\small{Questionnaire for qualitative evaluation (1 = strongly disagree, 5 = strongly agree). 
    %The questionnaire consisted of 12 questions divided into four categories: Presence, Embodiment, Reality, and Awareness.
    The results showing a significant difference (\textit{p} $<$ 0.05) are highlighted in bold in the table.}}
    \label{tb:questionnaire}
\end{table*}
\section{discussion}
\label{chap:discussion}

\subsection{Avoidance Motion}

\begin{figure}[t]
    \centering
    \subfigure[{\it NoHaptics}]{
        \includegraphics[width=0.45\hsize]{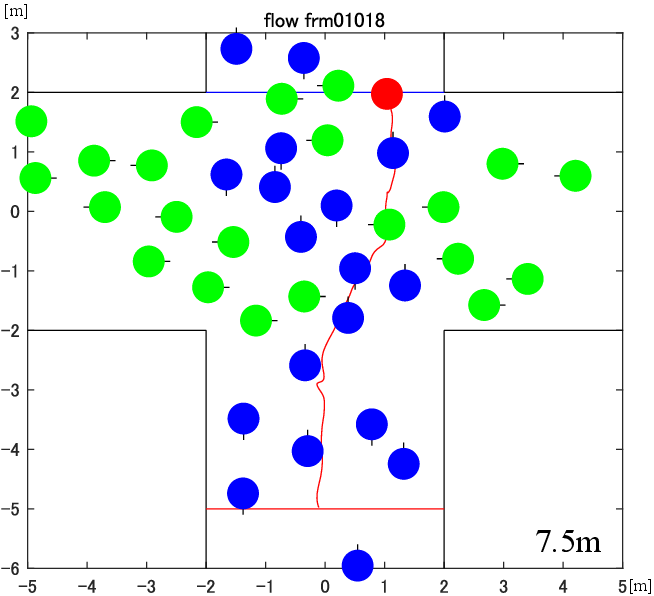}}
    \subfigure[{\it Haptics}]{
        \includegraphics[width=0.45\hsize]{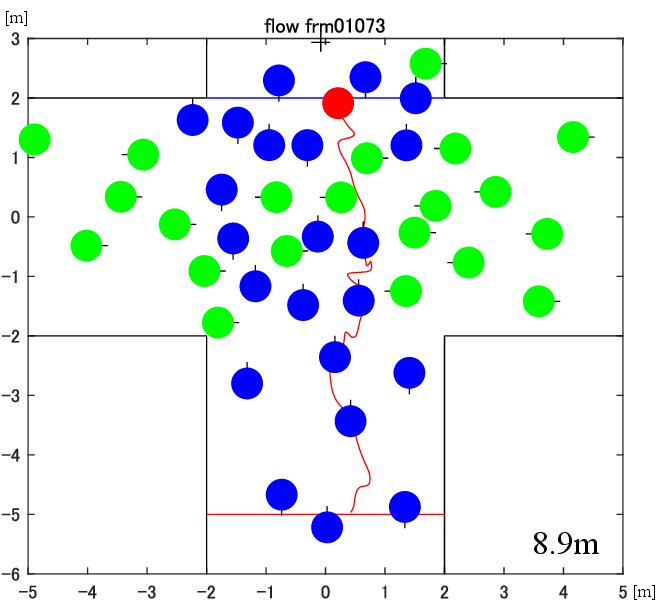}}
    \caption{\small{Examples of the walking trajectory of a subject (indicated by the red line and circle) in (a) {\it NoHaptics} and (b) {\it Haptics} cases.
    The blue and green circles indicate NPCs' positions.}}
    \label{fig:trajectory}
\end{figure}

Fig. \ref{fig:walking_motion}(a) and (b) shows that, the walking trajectory length and the pelvis angle displacement toward the goal direction increase in \textit{Haptics} compared to those of \textit{NoHaptics} ($p =0.00001$ in the trajectory length, and $p=0.005$ in the pelvis angle displacement). 
Fig. \ref{fig:trajectory} shows examples of the walking trajectory of a subject (indicated by the red line and circle) in (a) {\it NoHaptics} and (b) {\it Haptics} cases.
In the figure, the blue and green circles indicate the NPCs' positions, while the black line represents their walking directions.
It is observed that the trajectory in \textit{Haptics} is longer and more winding than \textit{NoHaptics}.

Examining collision avoidance behavior in detail, we observed that in \textit{Haptics} the forward displacement significantly decreased as shown in Fig. \ref{fig:collision_motion}(a) ($p=0.002$) whereas the lateral displacement increases as shown in Fig. \ref{fig:collision_motion}(b) ($p=0.006$), except for \textit{Side} and \textit{Back} cases.
Especially in the case that an NPC was approaching from {\it Front}, this result matches our intuition that a user avoids a collision mainly by moving in the lateral direction.
This result also suggests that the haptic feedback enhanced this avoidance movement. 
In Fig. \ref{fig:collision_motion}(c), it is observed that the pelvis angle displacement was increased in \textit{Haptics} ($p=0.003$), which implies that the haptic feedback enhanced the avoidance movement by rotating the body trunk.

These results do not support H1 but suggest that the haptic feedback changed subjects' behavior so that they tried harder to avoid collisions with other NPCs.
It is worth noting that significant differences are observed not only in the \textit{Out-view} but also in the \textit{In-view} cases. This suggests that the haptic feedback enhanced the avoidance movements even when an approaching NPC is inside the field of view.
We consider that there would be some cases in which such an NPC was not recognized by a subject.
Therefore, there is a possibility that the haptic feedback supported users' perception to such an unrecognized NPC.
%Based on these results, we conclude that the haptic device increases avoidance behavior in high-density and dynamic environments.

\subsection{Visual Exploration}
From Fig. \ref{fig:walking_motion}(c), the average head angle relative to the pelvis increased in \textit{Haptics} ($p=0.047$). Additionally, this metric also increased in the instantaneous response as shown in Fig. \ref{fig:collision_motion}(c) ($p=0.035$). This supports H2: the haptic feedback enhanced the visual exploration so that a subject was more likely to look around the surroundings.
%These findings suggest that the haptic device enhances subject's active visual exploration.

\subsection{Walking Speed}
From Fig. \ref{fig:walking_motion}(d), we found no significant difference in the average walking speed, which does not support H3 ($p=0.53$). However, Fig. \ref{fig:walking_motion}(e)  indicates that the standard deviation of the walking speed increased in \textit{Haptics} ($p=0.0004$). 
We consider this to be because a subject tended to stop or yield at the moment of detecting a collision, which makes the walking speed deviation larger.
This result suggests that the haptic device enhanced subject's sensitivity to collisions.

\subsection{Questionnaire}
The results in Table \ref{tb:questionnaire} indicate that subjects are more aware of NPCs approaching from the side and behind them with haptic feedback, supporting H2. While previous research \cite{Berton2022} did not find a significant difference in the sense of presence in VR, our study observed some minor differences in this aspect. This suggests that in dense and dynamic predestrian flows, the haptic feedback would enhance the sense of presence. The difference in Q3 also explain the difference in subjects' avoidance behavior which is also against supporting H1. Similarly to previous studies, we did not find a significant difference in the sense of embodiment in VR. Additionally, we observed no difference in the perceived realism of the simulation compared to real-world movements, which remains an area for future research.

\subsection{Number of Collisions}
From Fig. \ref{fig:walking_motion}(f), the number of collisions increased in \textit{Haptics} ($p=0.029$). 
We consider this is because increased avoidance movements in \textit{Haptics} made a subject more likely to collide with other NPCs in this kind of dense and dynamic pedestrian flow.
\section{Conclusion}
In this study, we investigated the effect of haptic feedback in dense and dynamic pedestrian flows in a VR environment. 
The results are summarized as follows:
\begin{enumerate}
    \item Haptic feedback increased avoidance movements as the walking trajectory length and pelvis angle displacement were significantly increased. Instantaneous response of collision avoidance was also significantly increased, even when an NPC was inside the field of view.
    This result suggests that the haptic feedback supported users' visual perception of NPCs.
    \item Haptic feedback enhanced visual exploration as the head angle relative to the pelvis significantly increased. This result suggested that subjects noticed NPCs approaching from the side and back, which was also observed in the questionnaire.
    \item Haptic feedback did not significantly change the walking speed in our experiment. However, its standard deviation increased, which implies that the haptic feedback enhanced subject's sensitivity to collisions.
\end{enumerate}


%%%%%%%%%%%%%%%%%%%%%%%%%%%%%%%%%%%%%%%%%%%%%%%%%%%%%%%%%%%%%%%%%%%%%%%%%%%%%%%%

%%%%%%%%%%%%%%%%%%%%%%%%%%%%%%%%%%%%%%%%%%%%%%%%%%%%%%%%%%%%%%%%%%%%%%%%%%%%%%%%

%%%%%%%%%%%%%%%%%%%%%%%%%%%%%%%%%%%%%%%%%%%%%%%%%%%%%%%%%%%%%%%%%%%%%%%%%%%%%%%%

%%%%%%%%%%%%%%%%%%%%%%%%%%%%%%%%%%%%%%%%%%%%%%%%%%%%%%%%%%%%%%%%%%%%%%%%%%%%%%%%

\bibliographystyle{bib/IEEEtran.bst}
\bibliography{bib/iros2025}

\end{document}